\documentclass[letterpaper, 10 pt, conference]{ieeeconf}  
\usepackage{amsmath,amssymb}
\usepackage{graphicx}
\usepackage{multirow}
\usepackage{epstopdf}
\usepackage{hyperref} 
\usepackage{fancyhdr}
\AppendGraphicsExtensions{.tif}
\IEEEoverridecommandlockouts
\title{\LARGE \bf
Closed-Loop Phase Selection in EEG-TMS Using Bayesian Optimization
}

\author{Miriam Kirchhoff$^{1,2}$, Dania Humaidan$^{1,2}$, Ulf Ziemann*$^{1,2}$
\thanks{$^{1}$ Department of Neurology \& Stroke, University of Tübingen, Tübingen, Germany}%
\thanks{$^{2}$ Hertie Institute for Clinical Brain Research, Tübingen, Germany}
\thanks{$^{*}$ {\tt\small ulf.ziemann@uni-tuebingen.de}}
}

\fancypagestyle{specialfooter}{%

  \fancyhf{}

  \fancyfoot[L]{\footnotesize This work was accepted by the International Conference on Systems, Man, and Cybernetics, Kuching, Malaysia, 2024. \\
  © 2024 IEEE. Personal use of this material is permitted. Permission from IEEE must be obtained for all other uses, in any current or future media, including reprinting/republishing this material for advertising or promotional purposes, creating new collective works, for resale or redistribution to servers or lists, or reuse of any copyrighted component of this work in other works.}
}

\begin{document}

\maketitle
\thispagestyle{empty}
\pagestyle{empty}

\begin{abstract}

Research on transcranial magnetic stimulation (TMS) combined with encephalography feedback (EEG-TMS) has shown that the phase of the sensorimotor mu rhythm is predictive of corticospinal excitability. Thus, if the subject-specific optimal phase is known, stimulation can be timed to be more efficient. In this paper, we present a closed-loop algorithm to determine the phase linked to the highest excitability with as few trials as possible. We used Bayesian optimization with different configurations as an automated, online search tool in an EEG-TMS simulation experiment. \\
From a sample of 38 healthy participants (25 f, 18 m), we selected all participants with a significant single-subject phase effect (N = 5) for the simulation. We then simulated 1000 experimental sessions per participant where we used Bayesian optimization to find the optimal phase. We tested two objective functions: Fitting a sinusoid in Bayesian linear regression or Gaussian Process regression. We additionally tested adaptive sampling using a knowledge gradient as the acquisition function compared with random sampling. We evaluated the algorithm’s performance in a fast optimization (100 trials) and a long-term optimization (1000 trials). \\
We found that for fast optimization, the Bayesian linear regression in combination with adaptive sampling gives the best results with a mean phase location accuracy of 79 \% after 100 trials. With either sampling approach, Bayesian linear regression performs better than Gaussian Process regression in the fast optimization. In the long-term optimization, Bayesian regression with random sampling shows the best trajectory, with a rather steep improvement and good final performance of 87 \% mean phase location accuracy. \\
In summary, we could show the suitability of closed-loop Bayesian optimization for phase selection. We show that we can increase the speed and accuracy by using prior knowledge about the expected function shape compared with traditional Bayesian optimization with Gaussian Process regression.

\end{abstract}


\thispagestyle{specialfooter}

\section{INTRODUCTION}

Transcranial magnetic stimulation (TMS) is a valuable treatment option for various neurological and mental disorders. For example, it has been shown that patients with neuropathic pain, stroke, and depression benefit from TMS treatment \cite{Lefaucheur2020}. Commonly, TMS is used to induce functional changes that last beyond the end of the stimulation session. Such effects are interpreted as neural plastic changes \cite{Huang2017}. \\
However, TMS shows high inter- and intra-subject variability \cite{Zarkowski2006}. One reason is that the neural states during stimulation fluctuate rapidly. These neural states can be predictive of the immediate effect of TMS \cite{Bergmann2018} and can be detected in electroencephalogram (EEG) data \cite{Tervo2020}. In the EEG data, the mu-band at the central motor cortex is commonly investigated \cite{Zrenner2022, Zrenner2023, Bergmann2018, Zarkowski2006}. Specifically, the phase of the mu-band has been shown to relate to neuronal firing. Higher neuronal firing rates were observed during the trough than at the peak \cite{Haegens2011}. Similarly, in TMS-EEG studies, corticospinal excitability was higher during the trough than at the peak \cite{Zrenner2022}. \\
In group analyses, there is a trend of the optimal phase, i.e., the one associated with the highest excitability, being in the trough \cite{Zrenner2022} or early rising phase \cite{Zrenner2023} of the mu-oscillation. However, for single-subject data, the optimal phase varies \cite{Torrecillos2020, Zrenner2023}. Thus, stimulating in the trough may improve TMS efficacy but is not necessarily the optimal temporal target. For individual subjects, it would be beneficial to find the optimal stimulation target phase with only few test pulses. A closed-loop algorithm could perform iterative improvement based on previously collected data, for example in preparation for an experiment or a clinical intervention. This paper thus aims to select the optimal phase of the mu-oscillation with as few trials as possible. \\
A similar challenge has been addressed by Tervo et al. \cite{Tervo2020}, optimizing coil orientation and location in TMS research, while minimizing the number of test pulses. To this end, they developed an algorithm that uses Bayesian optimization for an automated search. This search was iterative, where the optimal target location was updated after each measurement. Bayesian optimization also allows the use of an acquisition function, which informs the sampling process what the best next measurement target is \cite{Rasmussen2006, Frazier2018}. Thus, it is a suitable closed-loop method for research where samples are sparse. Additionally, Bayesian optimization does not require derivatives, making it flexible and computationally efficient \cite{Frazier2018}. It has been shown that this approach works for location and orientation search to increase motor evoked potentials (MEPs) \cite{Tervo2020} and TMS-evoked potentials (TEPs) \cite{Tervo2022}. On these grounds, we expect it is also possible to optimize the phase using Bayesian optimization. Since the relationship of phase and MEP is noisier than that of orientation or location and MEP, we expect that more trials will be required. In their paper, Tervo et al. \cite{Tervo2020} report a mean number of 18 trials required for convergence. In our work, we will attempt to optimize within 100 trials (5 min of stimulation with an inter-stimulus interval (ISI) of 3 s). We predict that using Bayesian optimization with an acquisition function will speed up the target search compared to a random search. \\
Tervo et al. \cite{Tervo2020} used Gaussian Process regression as their objective function, i.e., the function that is being optimized. This is a very common approach in Bayesian optimization \cite{Tervo2020, Rasmussen2006, Frazier2018, Tervo2022}. We want to additionally test Bayesian linear regression for fitting a sinusoid since prior literature has described the relationship between phase and MEP size to be sinusoidal \cite{Zrenner2020, Zrenner2022}. By using this prior knowledge, we can decrease the number of model parameters substantially. However, this model might perform badly at fitting non-sinusoidal single-subject data. Nevertheless, we expect it to outperform the Bayesian optimization using Gaussian Process regression. We will test both models using either random or adaptive sampling. \\
We used simulations to test Bayesian optimization for online phase selection. The data points were simulated based on real single-subject datasets that show a phase effect. All models were evaluated in terms of closeness to the optimal phase and to the best MEP size as well as their convergence behavior.

\section{METHODS}

\subsection{Experiment} 
\subsubsection{Participants}
Our sample consisted of 38 right-handed participants (25 female, 13 male) with a mean age of 25.7 years ($sd$ = 4.7). The participants did not have known medical conditions. The study was conducted in accordance with the declaration of Helsinki and was approved by the Ethics Committee of the Medical Faculty of the University of Tübingen (nr. 820/2021BO2). All participants were informed about the purpose of the study and signed informed consent.

\subsubsection{Procedure}
Participants’ handedness was determined using the Edinburgh Handedness Inventory \cite{Oldfield1971}. After preparing EEG caps and fixing EMG electrodes, we recorded 10 minutes of resting state EEG. \\
We conducted a hotspot search for the left motor cortex (M1), where a robust MEP in the abductor pollicis brevis (APB) or first dorsal interosseous (FDI) muscles could be observed. The resting motor threshold (RMT) was the minimal stimulation intensity that produced peak-to-peak amplitudes above 50 µV in at least half the test pulses. TMS pulses were administered using a MAG Pro R30 stimulator (MAG \& More, Munich, Germany) with a figure of eight coil (Cool-B65, inner winding diameter 35 mm). \\
In the main experiment, participants rested their hands and fixated on a constant visual target. The number of trials differed between two participant groups. The first group (N = 14) received one block of 1000 TMS pulses with an inter-stimulus-interval (ISI) of 2 s $\pm$ 0.25 s. The second group (N = 24) received four blocks of 300 TMS pulses (1200 total) with an ISI of 3 s $\pm$ 0.5 s. The TMS intensity was 110 \% RMT in both protocols.

\subsubsection{EMG and EEG recording}
Two EMG electrodes were fixed to the hand above the FDI and the right APB. The EMG data was recorded at 5 kHz. We used adhesive hydrogel electrodes in a bipolar belly-tendon montaging. \\
For the EEG recording, we used a 128 channel EEG-cap (EasyCap BC-TMS-128, EasyCap, Herrsching, German) with 10-5 electrode placement and a sampling frequency of 5 kHz. The reference electrode was placed on FCz, ground on AFz. To decrease movement artefacts and discomfort for participants, we fixated their heads with a vacuum pillow (Vacuform, Salzbergen, Germany). EMG and EEG recordings were amplified using a 24-bit NeurOne biosignal amplifier (Bittium, Oulu, Finland). 

\subsection{Experimental data processing}
We used Matlab 2024b with EEGlab 2024.0 \cite{Delorme2004} and FieldTrip 20231025 \cite{Oostenveld2011}. The script was uploaded to GitHub at \href{https://github.com/MiriamKirchhoff/BO_for_EEG-TMS_phase}{MiriamKirchhoff/BO\_for\_EEG-TMS\_phase}. 

\subsubsection{EMG processing}
We first baseline corrected and detrended the EMG data. Then, we calculated a principal component analysis (PCA) of the signals from FDI and APB for each participant. Both signals were then combined by projecting them onto the first coefficient of the PCA to maximize the explained variance of the signal (see \cite{Zrenner2023}). 
For the MEP calculation, the data were epoched between 20 and 40 ms w.r.t. TMS onset. We calculated the MEP as the peak-to-peak amplitude: First, we detected all peaks with a minimum prominence of one. We then calculated the maximum range between a positive and negative peak. Trials without peaks were rejected since no successful stimulation could be detected. The resulting MEPs were log-transformed and z-scored participant-wise for group-level statistical analyses. \\
For the rejection of trials with pre-innervation, we epoched the MEP data between -100 and -5 ms. After detrending, trials with a signal amplitude above 50 µV were rejected.

\subsubsection{EEG processing}
We epoched the EEG data in a range of -705 to -5 ms with respect to the TMS onset. The data was then baseline corrected and detrended. Afterwards, noisy EEG trials were rejected when one channel used in the spatial filter had a range $>$ 150 µV (see \cite{Zrenner2023}). The data was then downsampled to 1 kHz and bandpass filtered between 9 and 13 Hz (Hamming windowed sinc FIR filter). Afterwards, we filtered the data spatially using a Laplacian montage centered around C3 (C3 Hjorth filter). For the calculation of the mu-phase at the TMS onset, we used the PHASTIMATE function \cite{Zrenner2020} with standard settings. We also used their proposed way of calculating the phase estimation accuracy and calculated the signal-to-noise ratio (SNR) based on the resting state EEG data.

\subsubsection{Statistical testing}
We rejected trials based on EMG pre-innervation, EEG noise, and lack of EMG peaks. To test the group-level effect of phase on MEP size, we fitted a circular-to-linear model to the data with phase as a circular predictor variable. To test whether single subjects showed a phase effect, we repeated this procedure on single-subject data. To account for multiple comparisons, we corrected p-values using the Bonferroni procedure. Participants with a significant ($p < 0.05$) predictive effect of phase on MEP size were used for the simulation experiment.

\subsection{Simulation of Bayesian optimization for phase selection
}

\subsubsection{Bayesian optimization}
Bayesian optimization aims to optimize an unknown objective function $f$. To capture our knowledge about the function, we fit a prior function to the data. The shape of this prior function depends on the objective function selected. Evaluating the prior function at various potential measurement points gives the posterior distribution, which captures the expected shape given future measurements. From this posterior function, we can calculate an acquisition function, showing which point should be measured next given a certain aim. Our aim was to find the global maximum. We used a knowledge gradient \cite{Scott2011} as our acquisition function, as recommended by earlier work \cite{Tervo2020, Picheny2013}. This process is iterative and will be repeated until a stopping criterion is met. Stopping criteria can for example be a maximum number of iterations or, in our case, when the estimated location of the optimum did not change for several iterations.

\subsubsection{Objective functions}
In this study, we tested two objective functions: A Gaussian Process regression \cite{Rasmussen2006} and a Bayesian linear regression.

Gaussian Process regression. Using Gaussian Process regression for Bayesian optimization is very common \cite{Rasmussen2006, Frazier2018} and has been used in prior research on online EEG-TMS optimization \cite{Tervo2020, Tervo2022}. Using a Gaussian Process allows the objective function to take on diverse shapes, constrained by kernel choice. This can be advantageous since no function shape is assumed prior to the measurements. However, a good function fit may require more samples, especially for noisy data. We selected a periodic kernel $k$ with two free parameters: The length-scale $l$, determining the smoothness of the function, and the output variance $\sigma^2$.
$$
k = \sigma^2 \exp \left(- \frac{2 \cdot \sin(|x - x’|/2)^2}{l^2} \right) \eqno{(1)}
$$
The period was thus fixed to $2\pi$. The parameters were initialized at $l = \max_i \left(|y_i - \bar{y}|^2\right)$ and $\sigma^2$ as the variance of the initial sample $\mathbf{y}$ (see \cite{Tervo2020}).

Bayesian linear regression. Prior literature \cite{Torrecillos2020, Zrenner2022, Zrenner2023}, states that the phase effect follows a sinusoidal shape. Using this knowledge, we can minimize the number of degrees of freedom by only optimizing amplitude $A$, phase shift $\phi$, and intercept $I$. Thus, we optimized a standard sine function with a period of $2\pi$. This standard sine can be expressed as a combination of linear basis functions.
$$
f(x) = A(x+\phi) + I = w_0+w_1(x)+w_2(x) \eqno{(2)}
$$

By optimizing function $(2)$ in a Bayesian linear regression, we could calculate the posterior function. The posterior function was then used as input to the acquisition function.

\subsubsection{Acquisition function}
In this paper, we compared using random sampling to adaptive sampling using an acquisition function. For the latter, we used a knowledge gradient \cite{Scott2011}, as used in prior work \cite{Tervo2020, Tervo2022}, since it shows good performance in noisy optimization problems \cite{Picheny2013}. The knowledge gradient $KG(x_{N+1})$ shows the expected change of the maximum in relation to the location of the next datapoint to be sampled $x_{N+1}$. 
$$
KG(x_{N+1}) = \mathbb{E}[\mu^{\max}_{N+1}] - \mu^{\max}_ N. \eqno{(3)}
$$
In equation $(3)$, we express the maximum of the posterior mean function $ \mu^{\max}_{N+1} $ if we were to sample at location $x_{N+1}$. We sampled at the maximum of the acquisition function.

\subsubsection{Simulation of data points}
Data were simulated for each participant separately. We simulated the size of the MEP at a given phase. To this end, we used a sliding window centered around the phase of interest, covering 10 \% of the function domain. The real data falling within this window was used to fit a Gaussian distribution. From the resulting Gaussian distribution, we drew a random data point, which would be the simulated MEP size corresponding to the phase. After drawing 10 initial data points, we would use the random or adaptive sampling algorithm to determine the next target phase.

 \begin{figure}[thpb]
      \centering
      \framebox{
      \includegraphics[width=0.96\columnwidth]{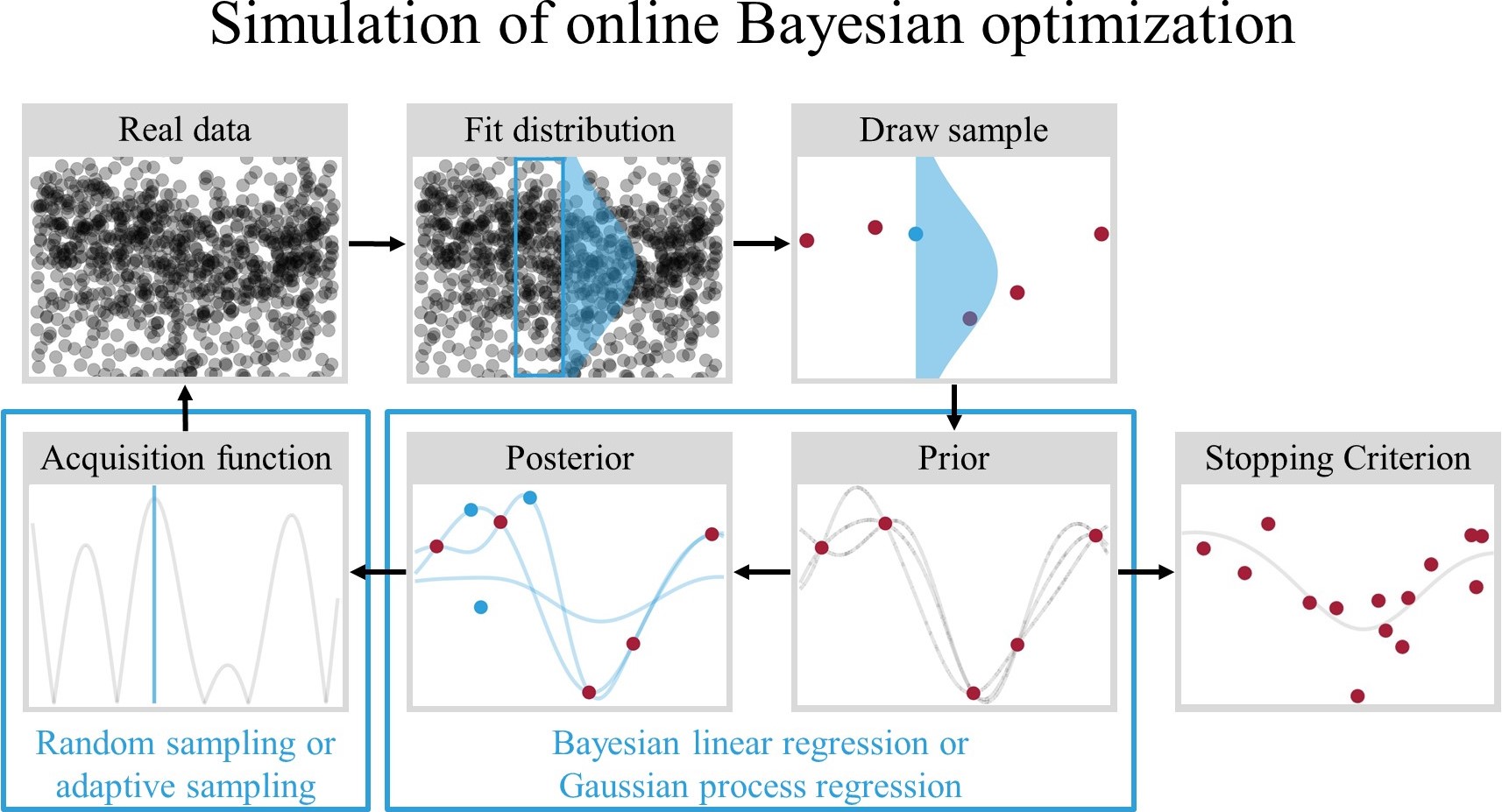}}
      \caption{\textbf{Simulation of online Bayesian optimization.} A datapoint at a selected location is simulated from real single-subject data by applying a sliding window, fitting a Gaussian distribution over the selected data, and drawing a random sample from it. Simulated data is used to fit a prior function using either Gaussian Process regression or Bayesian linear regression. If the stopping criterion is not met, a posterior function is calculated. An acquisition function determines the datapoint that maximizes the information gain for finding the optimal phase and determines the sampling location.}
      \label{figurelabel}
   \end{figure}

\subsubsection{Analysis of the simulation}
We first investigated the development of model accuracy over the number of samples. This was evaluated by calculating the relative distance of the ground truth optimal phase to the model-selected optimal phase at each iteration. We used the same non-parametric sliding window estimate as for the simulation to determine the optimal phase. We calculated the mean accuracy and 95 \% confidence interval for both objective functions (Gaussian Process regression and Bayesian linear regression) in combination with random sampling or an acquisition function.
Similarly, we also compared the corresponding size of the MEP at the selected phase with the MEP size at the optimal phase. Here, we calculated the accuracy as a percentage of the distance between the mean score and the optimal score.
We additionally calculated where the algorithm would have converged for different criteria. Convergence criteria ranged from 1 - 20 consecutive iterations where the maximum did not change by more than 5°. For these criteria, we calculated the iteration and accuracy.

\section{RESULTS}
We rejected 16.43 \% of trials. Of these, 0.62 \% were rejected based on a too high EEG threshold, 6.95 \% were rejected because no peak could be found in the EMG data, and 9.14 \% due to pre-innervation in the EMG.

\subsection{Phase effect}
\subsubsection{Phase estimation accuracy}
Using PHASTIMATE’s phase estimation accuracy, we could observe a mean phase estimation error of 3.28° ($sd$ = 57.08°). The data had a mean SNR of 2.67 dB ($sd = 13.13$  dB).

\subsubsection{Statistical analysis}
We first fitted a linear model with phase as a circular variable. At group level, we found that phase significantly predicts MEP size ($R^2 < 0.001$, $ F(2, 36606) = 13.10$, $ p < 0.001$). In the linear model, the phase for the highest MEP size was located at -149°, i.e., in the early rising phase (Fig. 2, left). At single-subject level, the most common optimal phase was also at the early rising phase (Fig. 2, right). We found that of the 38 subjects, 5 subjects show a significant phase effect ($p < 0.001$ for all) after Bonferroni correction. All these subjects had their optimal phase at the trough or early rising phase, between -172° and -128°. The effect sizes $R^2$ ranged between 0.022 and 0.035 ($\mu = 0.026$, $ sd = 0.005$).

 \begin{figure}[thpb]
      \centering
      \framebox{
      \includegraphics[width=0.96\columnwidth]{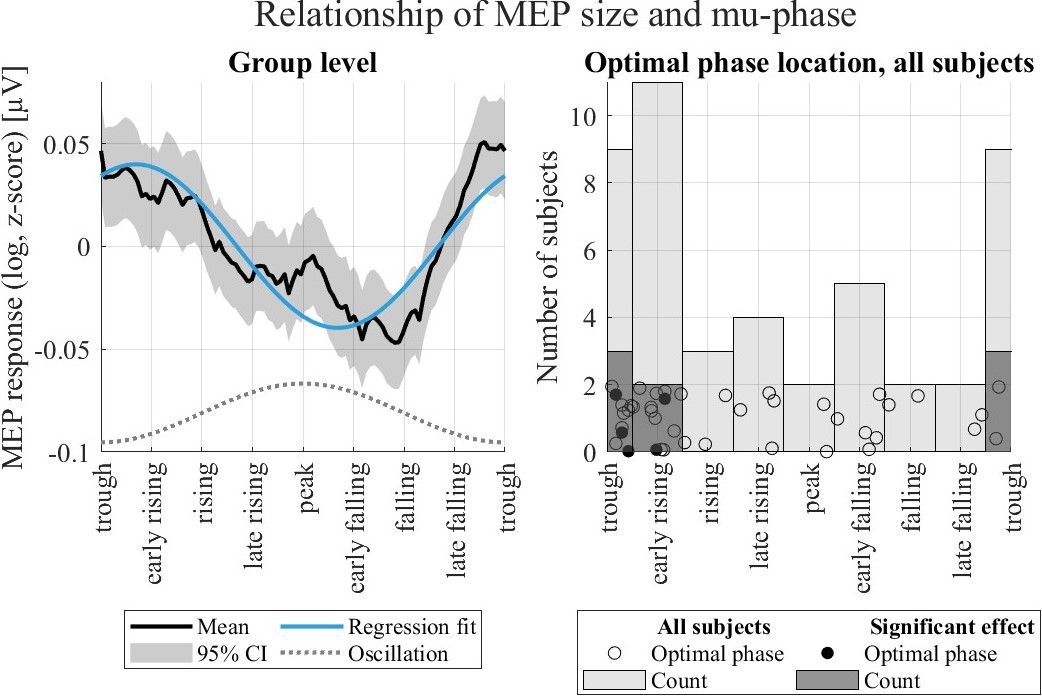}}
      \caption{\textbf{Left: Dependency of phase and MEP at group level.} The plot shows the sliding window mean $\pm$ 95 \% CI of the MEP size at different phases (black line and shaded area) as well as the regression fit (blue line). The highest MEPs were found in the negative peak. The lower gray line shows the corresponding phase oscillation. \textbf{Right: Optimal phase location for all subjects.} Dots and bars illustrate the location of the optimal phase for single subjects. Filled dots and dark gray bars show single subjects with a significant phase effect. }
      \label{figurelabel}
   \end{figure}

\subsection{Simulation results}
\subsubsection{Accuracy over number of samples}
The progression of accuracy (Fig. 3) shows that all models improve over time. At the starting point, all models show ca. 60 \% accuracy, i.e., ten random samples improve the phase estimate by 10 \% from random for both the MEP accuracy and the phase accuracy (see Table 1, 2). During the initial 100 samples, the phase location accuracy improves to 75 - 79 \% (Fig. 3, top left, Table 1). This improvement is highest for the Bayesian linear regression with adaptive sampling, followed by the same model with random sampling. Gaussian Processes perform almost equally in both sampling procedures. MEP size improved to 75 - 77 \% (Fig. 3, top right, Table 2), where Bayesian linear regression also performed better than Gaussian Process regression. The sampling approach did not influence this type of accuracy. \\
For long-term optimization, the phase accuracy improves to 85 - 87 \% (Fig. 3, bottom left, Table 1). Bayesian regression using an acquisition function seems to saturate fastest, showing little improvement in the phase accuracy above 600 trials with an accuracy of 85 \% (see Table 1) after 1000 trials. The other algorithms continue their improvement with the best final results by Bayesian regression using random sampling and Gaussian Processes using adaptive sampling. A similar pattern was found for the MEP accuracy which rises to 82 - 83 \% (Fig. 3, bottom right, Table 2). Long-term, the Bayesian regression using adaptive sampling shows the fastest-increasing accuracy score but converges early. The Bayesian regression using random sampling shows slightly slower decay but converges later.


\begin{table}[h]
\caption{Mu-phase location accuracy across iterations (µ $\pm$ 95 \% CI)
}
\begin{tabular}{|l|l|l|l|l|}
\hline
\textbf{\begin{tabular}[c]{@{}l@{}}Objective \\ function\end{tabular}} & \textbf{Sampling} & \textbf{\begin{tabular}[c]{@{}l@{}}Starting \\ (n = 0)\end{tabular}} & \textbf{\begin{tabular}[c]{@{}l@{}}Fast \\ (n = 100)\end{tabular}} & \textbf{\begin{tabular}[c]{@{}l@{}}Long-term \\ (n = 1000)\end{tabular}} \\ \hline
\multirow{3}{*}{\begin{tabular}[c]{@{}l@{}}Bayesian \\ linear \\ regression \end{tabular}}  & Adaptive          & \begin{tabular}[c]{@{}l@{}}60.8 \%\\ {[}59.1 62.5{]}\end{tabular}    & \begin{tabular}[c]{@{}l@{}}79.3 \%\\ {[}78.2 80.4{]}\end{tabular}  & \begin{tabular}[c]{@{}l@{}}85.3 \%\\ {[}84.7 85.8{]}\end{tabular}        \\ \cline{2-5} 
& Random            & \begin{tabular}[c]{@{}l@{}}60.2 \%\\ {[}58.4 61.9{]}\end{tabular}    & \begin{tabular}[c]{@{}l@{}}78.0 \%\\ {[}76.8 79.1{]}\end{tabular}  & \begin{tabular}[c]{@{}l@{}}86.9 \%\\ {[}86.2 87.5{]}\end{tabular}        \\ \hline
\multirow{3}{*}{\begin{tabular}[c]{@{}l@{}}Gaussian \\ Process \\ regression\end{tabular}} & Adaptive          & \begin{tabular}[c]{@{}l@{}}60.7 \%\\ {[}59.0 62.4{]}\end{tabular}    & \begin{tabular}[c]{@{}l@{}}75.2 \%\\ {[}73.9 76.6{]}\end{tabular}  & \begin{tabular}[c]{@{}l@{}}86.9 \%\\ {[}86.1 87.8{]}\end{tabular}        \\ \cline{2-5} 
& Random            & \begin{tabular}[c]{@{}l@{}}59.7 \%\\ {[}58.0 61.5{]}\end{tabular}    & \begin{tabular}[c]{@{}l@{}}75.9 \%\\ {[}74.7 77.1{]}\end{tabular}  & \begin{tabular}[c]{@{}l@{}}85.3 \%\\ {[}84.5 86.0{]}\end{tabular}        \\ \hline
\end{tabular}
\end{table}


\begin{table}[h]
\caption{MEP size accuracy across iterations (µ $\pm$ 95 \% CI)
}
\begin{tabular}{|l|l|l|l|l|}
\hline
\textbf{\begin{tabular}[c]{@{}l@{}}Objective \\ function\end{tabular}}           & \textbf{Sampling} & \textbf{\begin{tabular}[c]{@{}l@{}}Starting \\ (n = 0)\end{tabular}} & \textbf{\begin{tabular}[c]{@{}l@{}}Fast \\ (n = 100)\end{tabular}} & \textbf{\begin{tabular}[c]{@{}l@{}}Long-term \\ (n = 1000)\end{tabular}} \\ \hline
\multirow{3}{*}{\begin{tabular}[c]{@{}l@{}}Bayesian \\ linear \\ regression \end{tabular}}  & Adaptive          & \begin{tabular}[c]{@{}l@{}}61.1 \% \\ {[}59.4 62.8{]}\end{tabular}   & \begin{tabular}[c]{@{}l@{}}77.1 \% \\ {[}76.1 78.0{]}\end{tabular} & \begin{tabular}[c]{@{}l@{}}82.1 \% \\ {[}81.5 82.6{]}\end{tabular}       \\ \cline{2-5} 
& Random            & \begin{tabular}[c]{@{}l@{}}60.7 \%\\ {[}59.0 62.3{]}\end{tabular}    & \begin{tabular}[c]{@{}l@{}}77.0 \% \\ {[}76.0 78.0{]}\end{tabular} & \begin{tabular}[c]{@{}l@{}}82.7 \% \\ {[}82.1 83.2{]}\end{tabular}       \\ \hline
\multirow{3}{*}{\begin{tabular}[c]{@{}l@{}}Gaussian \\ Process \\ regression \end{tabular}} & Adaptive          & \begin{tabular}[c]{@{}l@{}}60.5 \% \\ {[}58.8 62.2{]}\end{tabular}   & \begin{tabular}[c]{@{}l@{}}75.1 \%\\ {[}74.0 76.2{]}\end{tabular}  & \begin{tabular}[c]{@{}l@{}}82.9 \% \\ {[}82.2 83.6{]}\end{tabular}       \\ \cline{2-5} 
& Random            & \begin{tabular}[c]{@{}l@{}}60.2 \% \\ {[}58.5 61.9{]}\end{tabular}   & \begin{tabular}[c]{@{}l@{}}75.7 \% \\ {[}74.7 76.7{]}\end{tabular} & \begin{tabular}[c]{@{}l@{}}82.0 \% \\ {[}81.4 82.6{]}\end{tabular}       \\ \hline
\end{tabular}
\end{table}

 \begin{figure}[thpb]
      \centering
      \framebox{
      \includegraphics[width=0.96\columnwidth]{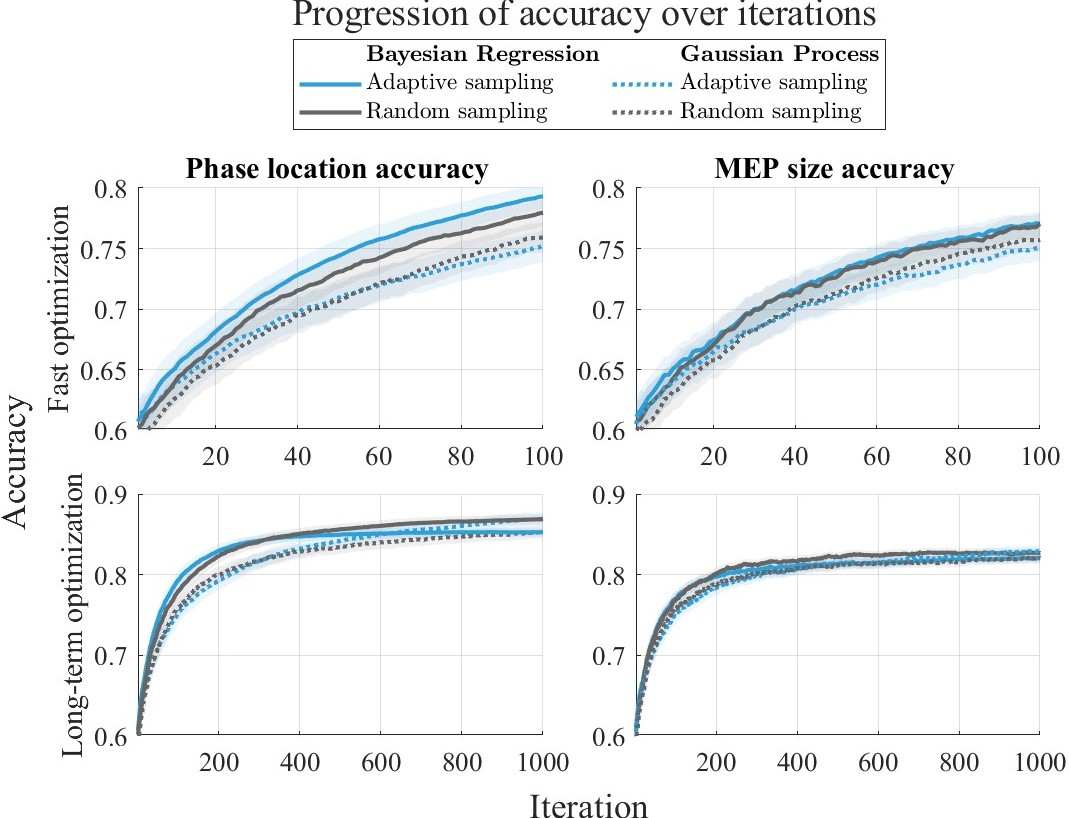}
}
      \caption{\textbf{Progression of accuracy across iterations.} Full lines for the Bayesian linear regression, dashed lines for Gaussian Process regression. Blue lines show adaptive sampling, gray lines show random sampling. Left column: Accuracy with respect to the phase location. Right column: Accuracy with respect to the MEP size. Top row: Fast optimization during initial 100 samples. Bottom row: Long-term optimization for 1000 samples.
}
      \label{figurelabel}
   \end{figure}

\subsubsection{Convergence criteria}
For convergence criteria of 1 - 20 consecutive samples that change the maximum by less than 5°, the accuracy and iteration of convergence are shown in Fig. 4. Bayesian regression using random samples shows the highest accuracy but also requires many more iterations to converge than the alternative approaches. Bayesian linear regression requires similar numbers of iterations to converge in either sampling approach but shows higher accuracy for adaptive sampling. Gaussian Processes using random sampling uses more samples than the adaptive sampling approaches though performs similarly to Bayesian regression using adaptive sampling. 

 \begin{figure}[thpb]
      \centering
      \framebox{
      \includegraphics[width=0.96\columnwidth]{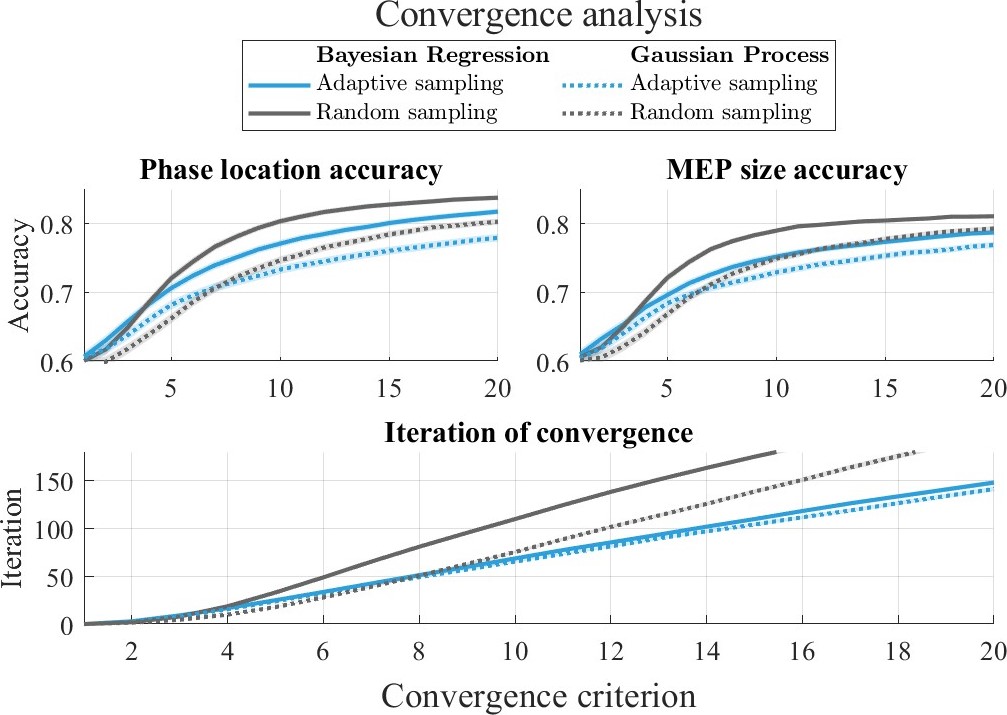}
}
      \caption{\textbf{Convergence analysis.} Full lines for the Bayesian linear regression, dashed lines for Gaussian Process regression. Blue lines show adaptive sampling, gray lines show random sampling. Top row: Phase location accuracy and MEP size accuracy at convergence are shown for different criteria. Bottom row: Iteration at which the different methods converge. 
}
      \label{figurelabel}
   \end{figure}

\section{DISCUSSION}
With this research, we illustrated the suitability of adaptive sampling algorithms using Bayesian optimization for online phase optimization in closed-loop EEG-TMS experiments. Overall, all tested approaches showed substantial improvement over time with varying slope and saturation points. For fast optimization, Bayesian optimization using Bayesian linear regression gives a superior phase estimate in terms of distance to the optimal phase. With regard to the size of the MEP, Bayesian regression performs very similarly when using adaptive or random sampling. For the fast optimization, Bayesian linear regression generally gives better accuracy than using Gaussian Process regression. For long-term optimization, the random sampling with Bayesian regression shows the best trajectory. \\
Our simulation showed phase and MEP accuracies of 60 \% at its initial point, indicating that the 10 initial random samples increase accuracies by 10 \% from random (50 \%). During the following 100 samples (fast optimization), the accuracy increases to 75 - 79 \%. This shows that generally, high accuracies can be found for small sample sizes already, with the best performance shown for adaptive sampling in a Bayesian linear regression. This method may be especially suited for small but noisy sample sizes since it strongly restricts the resulting function shape. This makes it more robust to noise and avoids overfitting. The acquisition function also exploits the maxima more strongly, increasing the accuracy early on but also converging earlier. Thus, using an acquisition function and prior knowledge about the expected sinusoidal shape of the function helps find a good fit with few samples. This goes along with our hypothesis. \\
During the following iterations of up to 1000 samples, this accuracy score was further increased to 85 - 87 \%. During this interval, the Bayesian linear regression shows the fastest improvement, followed by saturation. This saturation can be predicted to occur soon after 1000 trials for the random sampling and can be observed at ca. 600 samples for the adaptive sampling. This can be explained by the combination of the knowledge gradient acquisition function with the strongly pre-defined shape of the objective function with only three free parameters: After a certain number of samples, there is only little change in the function to be expected by measuring the next sample. Since the knowledge gradient expresses the expected change of the maximum, the algorithm selects points very close to the previous maximum, since distal points are unlikely to change the maximum at all. Thus, the algorithm shows no further improvement. Meanwhile, a random search gives more diverse information and may still change the shape after a high number of trials by recording many distal points.
For Gaussian Process regression, we can observe that a random search requires many samples to approximate the maximum well. Here, the adaptive sampling shows superior performance even in higher sample size, since more diverse shapes can be expected. Thus, a shifting of the maximum is more likely for distal points. At 1000 samples, a saturation can not yet be predicted. \\
These dynamics nicely illustrate that algorithm selection should be guided by the aim of the intervention: If a fast approximation is the goal, adaptive sampling with high constraints is beneficial, as shown in \cite{Tervo2020, Tervo2022}. On the contrary, for long-term optimization, adaptive sampling is less beneficial and may get stuck in local maxima. Higher degrees of freedom for the objective function will in the long-term be able to fit closer to the true trajectory while for a fast approximation, it hinders finding a good target.
Thus, in a clinical or research setting, we would select a fast, adaptive algorithm that is informed on the expected shape. Here, a few minutes are sufficient to improve the evoked MEPs significantly. Convergence criteria should be selected in accordance with the selected algorithm and as a trade-off of accuracy and duration. \\
To speed this process up even further, future research may want to investigate the possibility of adding prior knowledge about the expected location of the optimal phase. At group level, we found the highest MEPs at the early rising phase, similar to \cite{Zrenner2023}. The wide spread of optimal phases at single-subject level (Fig. 2) illustrates that there is no universally true optimal phase across participants. However, a majority can still be found around the trough. Informing the algorithm about this distribution may aid even faster optimization. \\
Interpreting our results, one should keep in mind that we only investigated subjects where a phase effect could be found. At a single-subject level, we found only 13 \% of subjects had a significantly strong influence of the mu-phase on MEP size. This is a lower rate than, e.g., in \cite{Zrenner2023}, who observed a single-subject phase effect in 29 \% of the participants. This can partially be attributed to the Bonferroni correction we used, as opposed to no correction. We intentionally selected a conservative correction method to ensure the exclusion of very weak or even false positive effects.\\
Future research might additionally investigate how one can predict if a subject shows a phase effect and aim to increase the fraction of these subjects. Simple pre-screening criteria could be high mu-bandpower and SNR to ensure a good mu-phase estimate. Both may also be improved by using individualized analysis pipelines including individualized frequency bands \cite{Donoghue2020} and spatial filters \cite{Nikulin2011}. These would cause better signal-to-noise ratios which could increase the number of subjects showing a phase effect and improve optimization. \\
In the future, one could test our algorithm in real-time experiments. To improve stimulation effects further, one could include more variables to optimize simultaneously (e.g., phase and coil orientation simultaneously). Limitations of our algorithm are that the knowledge gradient acquisition function becomes computationally infeasible in multi-dimensional settings. Expected Improvement offers a more scalable alternative \cite{Frazier2018}. Furthermore, the initial samples are assumed to be representative of the sample. For more unstable variables, more initial samples have to be collected.
It would be interesting to compare our approach to the performance of other machine learning approaches.
Additionally, our approach could be tested on TEPs. This would allow generalization to areas without MEP responses, like many clinical applications. \\ 
Here, we showed that the algorithm can increase TMS-induced MEP responses. This is especially useful in research or clinical applications where higher TMS responses are desired without increasing stimulation intensity.

\section{CONCLUSIONS}

In this paper, we demonstrate the suitability of closed-loop, online optimization of mu-phase in EEG-TMS. We show that within a few trials, Bayesian optimization can improve the accuracy of mu-phase location and MEP size prediction using various implementations. Specifically, we demonstrate that using prior knowledge about the sinusoidal relationship speeds up this process. Accuracy is best when combined with adaptive sampling for fast optimization and with random sampling for long-term optimization.

\section{CODE AVAILABILITY}
The code for this publication can be found on GitHub under \href{https://github.com/MiriamKirchhoff/BO_for_EEG-TMS_phase}{MiriamKirchhoff/BO\_for\_EEG-TMS\_phase}.

\section{ACKNOWLEDGEMENTS}
The authors thank Paul Bürkner for his valuable insights into Bayesian optimization. This study is part of the ConnectToBrain project that has received funding from the European Research Council (ERC) under the European Union’s Horizon 2020 research and innovation programme (Grant agreement No. 810377).

\addtolength{\textheight}{-12cm} 
\bibliographystyle{ieeetr}
\bibliography{bibliography}
\end{document}